# Spin wave based tunable switch between superconducting flux qubits

Shaojie Yuan[1]*, Chuanpu Liu[2]*, Jilei Chen[2]*, Song Liu[1], Jin Lan[5], Haiming Yu[2]†,

Jiansheng Wu[†1], Fei Yan[1], Man-Hong Yung[1], Jiang Xiao[3]†, Liang Jiang[4]†, Dapeng Yu[1]†

[1]Institute for Quantum Science and Engineering, and Department of Physics, Southern University of Science and Technology, Shenzhen 518055, China
[2]Fert Beijing Research Institute, School of Electronic and Information Engineering, BDBC, Beihang University, 100191 Beijing, China
[3]Department of Physics and State Key Laboratory of Surface Physics, Fudan University, Shanghai 200433, China [4]Pritzker School of Molecular Engineering, The University of Chicago, Chicago, Illinois 60637, USA
[5]Center of Joint Quantum Studies and Department of Physics, School of Science, Tianjin University, Tianjin 300350, China
* Equally contributed authors.

† Corresponding authors

**Quantum computing hardware has received world-wide attention and made considerable progress recently. YIG thin film have spin wave (magnon) modes with low dissipation and reliable control for quantum information processing. However, the coherent coupling between a quantum device and YIG thin film has yet been demonstrated. Here, we propose a scheme to achieve strong coupling between superconducting (SC) flux qubits and magnon modes in YIG thin film. Unlike the direct $\sqrt{N}$ enhancement factor in coupling to the Kittel mode or other spin ensembles, with N the total number of spins, an additional spatial-dependent phase factor needs to be considered when the qubits are magnetically coupled with the magnon modes of finite-wavelength. To avoid undesirable cancelation of coupling caused by the symmetrical boundary condition, a CoFeB thin layer is added to one side of the YIG thin film to break the symmetry. Our numerical simulation demonstrates avoided crossing and coherent transfer of quantum information between the flux qubit and the standing spin waves in YIG thin films. We show that the YIG thin film can be used as a tunable switch between two flux qubits, which have modified shape with small direct inductive coupling between them. Our results manifest that it is possible to couple flux qubits while suppressing undesirable cross-talk.**

Quantum computing and simulation based on superconducting qubits have achieved significant progress in recent years (*1-3*). Many efforts were devoted to hybridizing the solid-state qubits with other physical systems, such as mechanical or magnetic systems (*4-9*). For instance, the Kittel mode of a macroscopic YIG sphere was coherently coupled to a transmon qubit in a 3D cavity with the microwave photons manipulated inside the cavity (*8*). Besides, the superconducting flux qubit was successfully hybridized with spin ensembles, i.e., nitrogen-vacancy (NV) centers in diamond via magnetic interaction (*4-6*). On the other hand, because of the zero Joule heating, the wave nature with microwave working frequency, spin wave (whose quanta is called magnon) has become a promising candidate for conventional information transmission and processing and acquired the potential to establish a spin-wave based computing technology, far beyond its CMOS counterpart (*10-16*). Due to its favorably low damping, ferrimagnetic insulator yttrium iron garnet (YIG) is particularly promising for these applications (*17-19*).

In this work, we propose a novel hybrid system, consisting of superconducting flux qubits and the standing spin waves (*20*) in ferrimagnetic YIG thin film. The latter system has been widely used in spintronics and magnonics (*17-19*), while, its magnetic coupling to superconducting qubits and the corresponding application in quantum information processing has not been extensively investigated. As shown in the following, unlike the coupling to spin ensembles or Kittel mode of spin waves (*4, 5, 7*), the enhancement factor for the coupling strength does not follow the $\sqrt{N}$ law, but carries a modulation associated with the finite spin-wave wavelength. In our proposal, an additional thin pinning layer of CoFeB is deposited on one side of the YIG thin film to break the symmetry at the boundary conditions (*21-30*). Avoided crossing of the energy spectrum can be numerically simulated by solving Heisenberg equation based on the full Hamiltonian of the flux qubit, the spin waves in the YIG thin film and their coupling. We find that it is possible to transfer quantum information coherently between the flux qubit and the spin wave mode in the YIG thin film. Moreover, we propose an experimentally feasible design to switch "on" and "off" the coupling between two shape-modified flux qubits or to entangle them via the perpendicular standing spin waves (PSSWS) of the YIG thin film. Hybridizing one flux qubit with and further "tuning" the inductive coupling, which causes cross-talk (*31*), between multiple flux qubits or entangling them through PSSWs highlights the application of spin wave bus in quantum computing, further expanding the application of spin wave-based computation technology (*32-35*).

A superconducting (SC) loop with three Josephson junctions compose of a flux qubit with the superposition of the clockwise and counter clockwise persistent currents state as the qubit ground state: $|g> = |↻> - |↺>$ and first excited state, $|e> = |↻> + |↺>$ (*36, 37*), respectively. The net currents and the resulting the magnetic field threading the loop for the $|g>$ and $|e>$ states are distinct. Consequently, the Rabi oscillation between the two states of the flux qubit generates an alternating magnetic field perpendicular to the SC loop, which can be used to excite spin waves in YIG system. The basic setup of the hybrid system is shown schematically in Fig. 1, which consists of a 5x 5 µm² superconducting loop and a 3x 0.08x 3 µm³ YIG thin film above. A much thinner CoFeB capping layer ~ 10 nm in thickness is deposited on the top side of the YIG thin film to pin the magnetization in YIG at the interface. The magnetization follows Dirichlet boundary condition at the pinned surface, and Neumann boundary condition at the other free surface (*21-27*). The resonant frequencies of the perpendicular standing spin wave (PSSW) modes are (*20*),

$$f_{pssw} = \frac{\gamma \mu_0}{2\pi} \sqrt{\left[H_{ext} + \frac{2A_{ex}}{\mu_0 M_s}\left(\frac{n\pi}{\delta}\right)^2\right]\left[H_{ext} + \frac{2A_{ex}}{\mu_0 M_s}\left(\frac{n\pi}{\delta}\right)^2 + M_s\right]} \quad (1)$$

with gyromagnetic ratio $\frac{\gamma}{2\pi} = 28$ GHz/T, vacuum permeability $\mu_0 = 1.256 * 10^{-6}$ N/A², saturation magnetization $M_s = 192$ kA/m for YIG, thickness $\delta = 80$ nm, the exchange constant $A_{ex} = 3.1$ pJ/m, external field $H_{ext}$ and mode number $n = 1,2,3,...,$ . Values of $M_s$ and $A_{ex}$ are obtained by fitting the resonance of a 295 nm YIG thin film from Ref (*19*) using equation (1) with mode number n = 1,2,3,4,5,6. Experiment has measured the resonance value for the PSSW mode of 80 nm YIG thin film at near zero external field to be 4.57 GHz, which is different from theoretical prediction 3.39 GHz. The discrepancy may be due to choosing of order parameter to be integer for unsymmetrical pinning in the fitting process, as actually there are ¾ wavelength in thickness direction for n = 1 as illustrated in Fig. 1 b. For our quantum control schemes, we will use the experimental resonance values and design a flux qubit with transition frequency close to $f_{pssw}^{(n=1)}$ and sufficiently detuned from the CoFeB resonance. Using the geometric confinement, the proper boundary conditions and the suitable coupling strength (see Eq. 4 with later discussion), the PSSW of wavelength of $\lambda = \frac{4}{3}\delta = \frac{4 \times 80}{3}$ nm can be excited. An external field of 10 Gauss is applied to align spins in YIG and the field created by the YIG thin film and the CoFeB capping layer on the flux qubit is of the same order (see Fig 1. c), assuming the spin density $n_{CoFeB} = 1.61 \times 10^{29}\ m^{-3}$ for CoFeB and $n_{YIG} = 2.14 \times 10^{28}\ m^{-3}$ for YIG. The distance between the flux qubit and the YIG thin film is chosen to be around 1-1.5 μm for later simulation in Fig.3. At these distances, the total magnetic field on the qubit is between 21.5 to 37 Gauss, which is less than the critical field of the aluminum superconductor (around 100 Gauss) and guarantees superconductivity of the flux qubit. In addition, other superconducting material such as Niobium can be used to fabricate the loop and junctions of the flux qubit, which has a much higher critical magnetic field for superconductivity, i.e., above 1000 Gauss.

From Ref (19), the decay rates for YIG thin film and CoFeB pinning layer are estimated as $\Gamma_{YIG,n=2} \sim 40$ MHz and $\Gamma_{CoFeB} \sim 300$ MHz, where n is the PSSW mode number. Since the decay rate is proportional to the frequency and the frequency is approximately proportional to the square of mode number, intrinsic decay rate for n=1 PSSW mode is ~ 10 MHz. The resonance frequency for n=1 PSSW mode in YIG thin film and CoFeB pinning layer are $f_{YIG} \sim 4.6$ GHz and $f_{CoFeB} \sim 1.35$ GHz and the exchange coupling strength is $g_{CoFeB,YIG} \sim 500$ MHz, which makes converted decay rate of CoFeB on n=1 PSSW mode as $\Gamma_{CoFeB \to YIG, n=1} = (\frac{g_{CoFeB-YIG}}{f_{YIG}-f_{CoFeB}})^2 * 300 \sim 7$ MHz and total decay rate for n=1 PSSW mode being 17 to 20 MHz. In our proposal, we replace the microwave antenna in Ref (19) a flux qubit loop, which has a lower decay rate and a much smaller inductive coupling with the sample, and we expect the magnon decay rates will be further reduced. Therefore, it is reasonable to assume that the decay rate for n=1 PSSW mode is about 20-30 MHz.

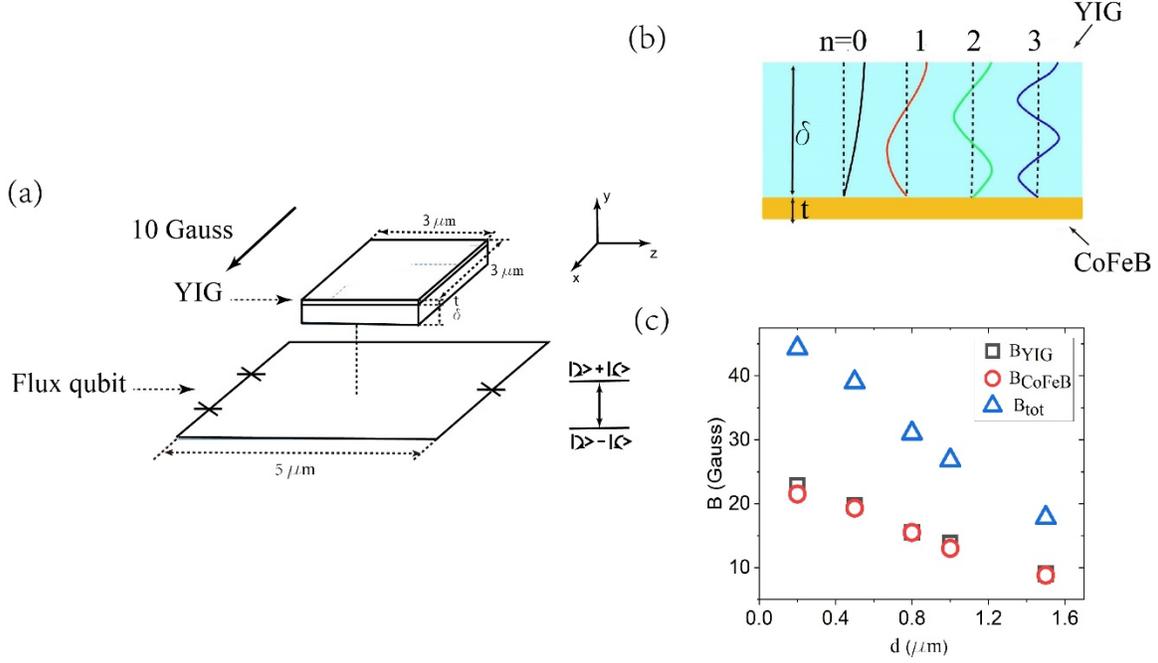

Fig. 1 **Hybrid structure coupling a flux qubit and a YIG thin film (spin wave, or a magnon).** (a) A YIG thin film with the dimension of 3 x 0.08 x 3 μm³ is placed in the center of the 5 x 5 μm² flux qubit loop separated by a distance d. An external field of 10 Gauss is applied along the x axis to align the spins in the YIG thin film. The thickness of YIG thin film is $\delta_{YIG}$=80 nm. A perpendicular standing spin wave (n = 1) with wavelength $\lambda = \frac{4}{3}\delta$ is excited. The frequency of the flux qubit is chosen to match that of the spin wave, which is experimentally around 4.7 GHz. The alternating magnetic field in the flux qubit loop excites the spin wave in the YIG thin film. (b) A cartoon depicts the PSSW of mode number n=0,1,2,3 with bottom spins being pinned and top spins unpinned. In our proposal, n=1 mode is selected. (c) The magnetic field on the flux qubit created by the YIG thin film and CoFeB thin layer. The spins in both YIG and CoFeB are fully aligned along x.

In the following, we consider the coupling strength between the flux qubit and YIG thin film. Hamiltonian for a ferromagnetic or ferrimagnetic material in a magnetic field takes the following general form (38)

$$\hat{H} = -\sum_{m,n} \hat{S}_m \cdot J_{m,n} \cdot \hat{S}_n + \sum_m \hat{S}_m \cdot B \quad (2)$$

with coupling matrix $J_{m,n}$ between the spins, with the assumptions that deviations from the group state are small, we can perform the Holstein–Primakoff approximation and transforming into the momentum space, we obtain

$$\hat{H} = -JNS^2 + NS\mu_B B_z + \sum_{\mathbf{k}}^{B.Z.} \sum_{i=1}^{3}(\hbar\omega_{k_i} + \mu_B \cdot B_z)a_{\mathbf{k}}^\dagger a_{\mathbf{k}} +$$

$$\sum_{\mathbf{k}}^{B.Z.} a_{\mathbf{k}}^\dagger \sum_{\mathbf{m}}(2S)^{\frac{1}{2}}\mu_B\left(\frac{1}{\sqrt{N}}e^{i\mathbf{k}.\mathbf{m}}\left(\frac{B_x+iB_y}{2}\right)\right) + a_{\mathbf{k}}\sum_{\mathbf{m}}(2S)^{\frac{1}{2}}\mu_B\left(\frac{1}{\sqrt{N}}e^{-i\mathbf{k}.\mathbf{m}}\left(\frac{B_x-iB_y}{2}\right)\right) \quad (3)$$

Where $i = 1,2,3$, $\hbar\omega_{k_i} = 2JS(1-\cos k_i) = 4JS\sin^2\left(\frac{k_i}{2}\right)$, $S$ is total spin at each lattice site, $\vec{k}$ is the wavevector of the spin wave and $N$ is the number of lattice sites in each direction. From Equation (3), by replacing the summation over each site with integration over space and insert the spin density, the integral form of the coupling strength between the flux qubit and YIG thin film is obtained as following:

$$g_{eff}^{\vec{k}} \sim \frac{(2S)^{\frac{1}{2}}\int \rho\mu_B \cdot e^{i\vec{k}.\vec{r}}\left(\frac{B_x+iB_y}{2}\right)d^3r}{(\int \rho d^3r)^{\frac{1}{2}}} \quad (4)$$

where $B(x,y,z)$ is the microwave excitation field created by the flux qubit and $\rho$ is the spin density in YIG. Unlike the simple $\sqrt{N}$ enhancement associated with coupling to Kittel mode, there is an extra spatial-dependent phase factor $e^{i\vec{k}.\vec{r}}$ in Eq. (4). For long wavelength spin wave, $|\mathbf{k}| \ll 1\ \mu m^{-1}$ and $e^{i\mathbf{k}.\mathbf{r}} \sim 1$ and if $B_x$ or $B_y$ only vary slowly compared to $\frac{1}{|\mathbf{k}|}$ in real space, $g_{eff}^{\mathbf{k}}$ will be proportional to $\sqrt{N}$. However, for the short wavelength spin wave, $g_{eff}^{\mathbf{k}}$ is not necessarily proportional to $\sqrt{N}$ and can even be zero if the integration region covers exactly integer times of the wavelength along wavevector direction. This is also the reason, to excite PSSW mode in YIG thin film by an almost homogenous field, an asymmetric boundary condition is required to avoid zero coupling strength caused by the phase factor.

Given the dimension of the flux qubit square loop 5x 5 $\mu m^2$ and the persistent current I ~ 500 nA (*39, 40*), the magnetic field produced by the flux qubit can be evaluated using Ampere's law $B(\mathrm{r}) = \frac{\mu_0}{4\pi}\oint I\frac{\overrightarrow{dl}\times\vec{r'}}{|\vec{r'}|^3}$ and $B_y$ dominates while, $B_x$, $B_z$ is close to zero in Fig .1. Given a net spin density $\rho = 2.14\times 10^{28}\ m^{-3}$ in YIG, we obtain the absolute value $g_{eff}^k$ as a function of the separation distance d between of the coupling strength between the flux qubits and the YIG thin film as in Fig. 2:

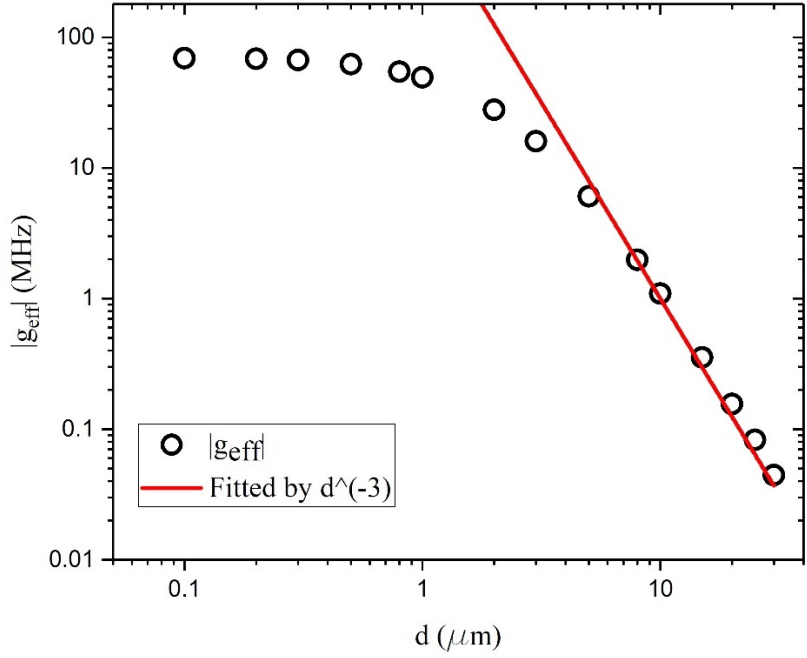

Fig. 2, **Coupling strength $|g_{eff}^k|$ as a function of the separating spacing d.** For small distance (d<2 μm), $|g_{eff}^k|$ for $\vec{k}_y = \frac{2\pi}{\lambda}$, where y is the direction in Fig. 1, decreases slowly with the distance and is above 30 MHz, which is larger than decay rate of magnon in YIG thin film. For large d, $|g_{eff}^k|$ decreases as $d^{-3}$ indicated by the red curve.

With the coupling strength estimated above, the full Hamiltonian with the flux qubit and YIG thin film can be written as

$$\hat{H} = -JNS^2 + NS\mu_B B_z + \sum_{\mathbf{k}}^{B.Z.}(\hbar\omega_{\mathbf{k}} + \mu_B \cdot B)a_{\mathbf{k}}^+ a_{\mathbf{k}} + \frac{h}{2}(\Delta\sigma_x + \varepsilon\sigma_z) - h(g_{eff}^k a_{\mathbf{k}}^\dagger + g_{eff}^{k*} a_{\mathbf{k}})\sigma_z + h\lambda\cos\omega t \cdot \sigma_z$$

(5)

where the first and second terms are the ferromagnetic and Zeeman terms, the third term describes the spin wave excitation, the fourth term is the flux qubit with $\Delta$ the tunneling energy splitting and $\varepsilon$ being the energy bias between the two qubit states, the fifth term characterizes interaction between the two devices, and the last term is the external driving of the flux qubit. Here, $\sigma_{x,z}$ are the Pauli matrices. The first two terms can be neglected for the reason that spin wave energy is a small perturbation compared to these two energies. By changing the basis of the flux-qubit, neglecting the Zeeman splitting and performing the rotating wave approximation, the Hamiltonian becomes,

$$\widehat{H} = \frac{h}{2}(\sqrt{\Delta^2 + \varepsilon^2} - \omega)\sigma_z + (\hbar\omega_k + \mu_B.B)a_k^+ a_k + h\frac{\Delta}{\sqrt{\Delta^2+\varepsilon^2}}\left(g_{eff}^k a_k^+ \sigma_- + g_{eff}^{k*} a_k \sigma_+\right) + h\frac{\Delta}{\sqrt{\Delta^2+\varepsilon^2}}\frac{\lambda}{2}(\sigma_+ + \sigma_-) \tag{6}$$

Where $\sigma_+, \sigma_-$ are rasing and lowering operator. $\sigma_z = 2*(\sigma_+\sigma_- - \frac{1}{2})$. Approximating the flux qubit as a harmonic oscillator and let $\sigma_+ \to \hat{c}^+$ and $\sigma_- \to \hat{c}$, the Hamiltonian can be written in a different form. Employing the Heisenberg relation $\frac{d\hat{c}}{dt} = [\hat{c}, H]$, sloving in Fourier space and transforming back to the lab frame, we obtain simulation of the energy spectrum

$$\tilde{\sigma}_{-,\omega} \sim \frac{1}{\omega-\sqrt{\Delta^2+\varepsilon^2}+i\Gamma_{fq}-\left|g_{eff}^k\frac{\Delta}{\sqrt{\Delta^2+\varepsilon^2}}\right|^2/(\omega-\omega_{sw}+i\Gamma_{sw})} \tag{7}$$

with $\omega$ being the driving pulse frequency and $\omega_{sw}$ being the resonance frequency of the standing spin wave of the YIG thin film. The expression of Eq. (7) describes the spectroscopic measurement of the flux qubit hybridized with spin waves in YIG thin film. Chossing paremters as $\frac{\Delta}{2\pi}$=4.52 GHz, $\frac{\Gamma_{FQ}}{2\pi} = 2$ MHz, $\frac{\omega_{sw}}{2\pi}$ =4.57 GHz and $\frac{\Gamma_{sw}}{2\pi} = 20$ MHz, which is a rsonable number since the decay rate for Kittel spin wave in a perfect sphere is around 1 MHz (7) and for finite wavelength spin wave in the YIG thin film is 6.8 MHz at 20 mk with GGG substrate and 1.4 MHz without substrate (41), let $\frac{|g_{eff}^k|}{2\pi} = 0$ MHz and 30 MHz, we obtain a simulated spectrum for a bare qubit and a hybridized qubit-spin wave system, as shown in Fig.3. The avoided cross or gap shows the strong coupling between flux qubit and standing spin wave of YIG thin film with vacuum Rabi splitting 2g = 60 MHz, which supports coherent energy or information exchange between them. Before preceding further, let us have a brief discussion about the influence of the CoFeB thin layer on the flux qubit. Using Eq. 4, with long wavelength approximation ($k\sim 0$) and spin density of CoFeB being $1.61 * 10^{29}$ $m^{-3}$ (Co), and $d = 1.2$ $\mu m$ as the parameter chosen in Fig. 3, a rough coupling strength between flux qubit and Co thin layer is 200 MHz. Decay rate for CoFeB is $\Gamma_{CoFeB} \sim 300$ MHz and the converted influence on the flux qubit from Co electrons would be $\Gamma_{Co} * \left(\frac{g}{\Delta}\right)^2 \sim 1.2$ MHz, where $g$ is the coupling strength and $\Delta$ is the off resonance between the flux qubit and CoFeB. We may introduce the the damping constant $\alpha = \frac{\Gamma}{f}$, where $\Gamma$ is the decay rate and $f$ is the resonace frequency. For YIG, $\alpha$ is on the order of $10^{-5}$ to $10^{-4}$, which makes decay rate as small as 3.3 MHz at a resonace of 4.57 GHz, most possibly by improving the thin film growing quality. In addition, a low ferromagntic alloy $Co_{25}Fe_{75}$ with damping constant as low as $5*10^{-4}$ is reoprted. This material could substitute the CoFeB capping layer, which would have the decay rate $\Gamma_{Co25Fe75} < 1$ MHz instead of $\Gamma_{CoFeB} \sim 300$ MHz and decrease the total decay rate of YIG-pinning layer to below 5 MHz. These further ensure the possibilities to implement thicknees mode of YIG thin film in quantum information processing.

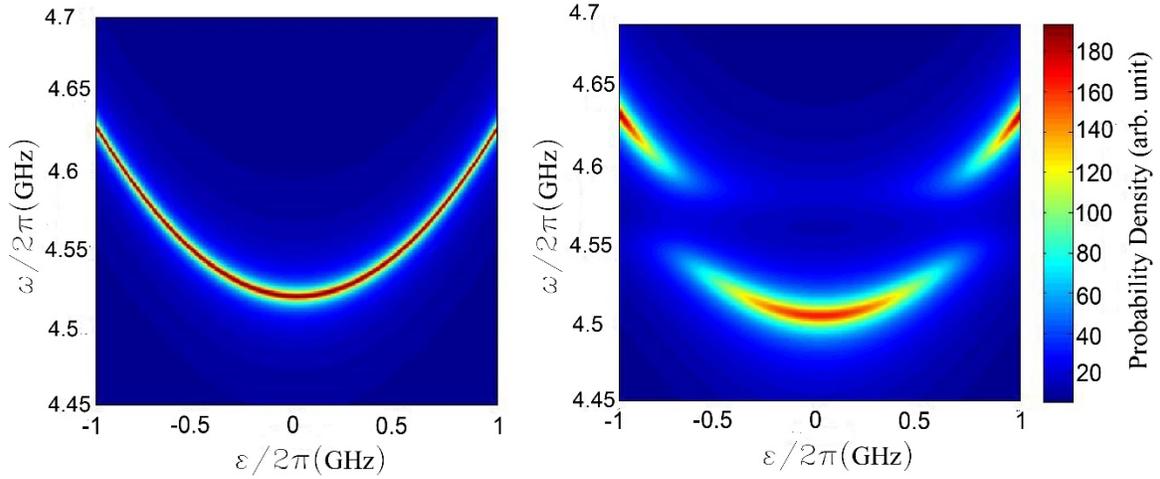

Fig. 3 **Simulation of the energy spectrum of a flux qubit coupled to standing spin waves in the YIG thin film.** (a) Spectrum of a bare flux qubit with $\Delta$=4.52 GHz, $\Gamma_{FQ} = 2$ MHz and $g_{eff}^k = 0$ in Eq. (7). (b) Spectrum of a flux qubit coupled to the standing spin wave of the flux qubit with $\frac{|g_{eff}^k|}{2\pi} = 30$ MHz, $\frac{\omega_{sw}}{2\pi} = 4.57$ GHz, $\frac{\Gamma_{sw}}{2\pi} = 20$ MHz.

Next, we propose a scheme to entangle and further switch the coupling "on" and "off" between two shape-modified flux qubits through PSSW mode in YIG thin film. Fig. 4 shows the schematic: two modified flux qubits with center-to-center distance of $20\sqrt{2}$ $\mu m$, are placed on top of a YIG thin film with a vertical separation d. The left/right arc of a flux qubit is a quarter of a 10 $\mu m$ radius circle and the top/down arc is a quarter of a 13.2 $\mu m$ radius circle. Mutual inductance of the two loops is given by the Neumann formula $L_{m,n} = \frac{\mu_0}{4\pi} \oint \oint \frac{dX_m dX_n}{|X_m - X_n|}$. The designed orientations of those arcs are to decrease the mutual inductance between the two flux qubit loops from several tens MHz for comparable size square loops to 3.97 MHz for the current design with circulating current as much as 500 nA. YIG thin film is ~ 80 nm in thickness with left/right sides being a quarter of a 10 $\mu m$ circle and top/down sides having the length of $10\sqrt{2}$ $\mu m$, which is also deposited with 10 $nm$ CoFeB on one side. As oscillation occurs between the two states of a flux qubit, alternating magnetic fields are created outside the loop and Fig. 4 (d) shows the coupling strength between each flux qubit and the YIG thin film as a function of the distance d in between. As shown in Fig. 4, stray magnetic field created by the YIG-CoFeB thin film is below the superconducting critical field of material of Niobium, i.e., 1000 Gauss, that is used to fabricate the flux qubit.

Readout of a flux qubit can be realized via another shaped-modified squid loop as in Fig. 4 c. Mutual inductance between the squid loop and flux qubit is $3.8 * 10^{-11}$ $H$, while the one between the squid loop and the neighboring flux qubit is $5.6 * 10^{-14}$ $H$. This guarantees that reading-out flux qubit will not be influenced much by the state of neighboring qubit, even operating simultaneously. Microwave line which is not shown, can quickly tune flux quit resonance frequency to the frequency of the (PSSW) spin wave mode of 4.57 GHz. At distance $d = 0.5$ $\mu m$, the absolute value of coupling strength is about 50 MHz. If both flux qubits are detuned simultaneously to 630 MHz below 4.57 GHz, effective coupling strength J between the two flux qubits can be

$$J \sim \frac{g_1 g_2 \left(\frac{1}{\Delta_1} + \frac{1}{\Delta_2}\right)}{2} \approx -3.97 \text{ MHz} \qquad (8)$$

This will cancel the mutual inductive coupling between (+ 3.97 MHz) the two flux qubits loops, thus switching off the coupling. On the other hand, if detuning both flux qubit to 400 MHz above 4.57 GHz, J would be 6.25 MHz, and plus additional mutual inductive 3.97 MHz, the total coupling strength would be about 10 MHz. Since the intrinsic life time for flux qubit can be about 1 μs, coupling strength of 10 MHz is strong enough to entangle the two qubits. In this way the coupling between two flux qubits is switched "on" and "off".
In addition, the intrinsic decay rate of thickness mode spin wave in YIG thin film is about
$\Gamma_{YIG}$ =10 MHz, which will introduce an extra broadening of $10*\left(\frac{50}{400}\right)^2 = 0.15$ MHz on
the flux qubit. Similarly, the CoFeB thin layer gives rise to another $300*\left(\frac{20}{3300}\right)^2 \sim 0.01$
MHz broadening on flux qubit.

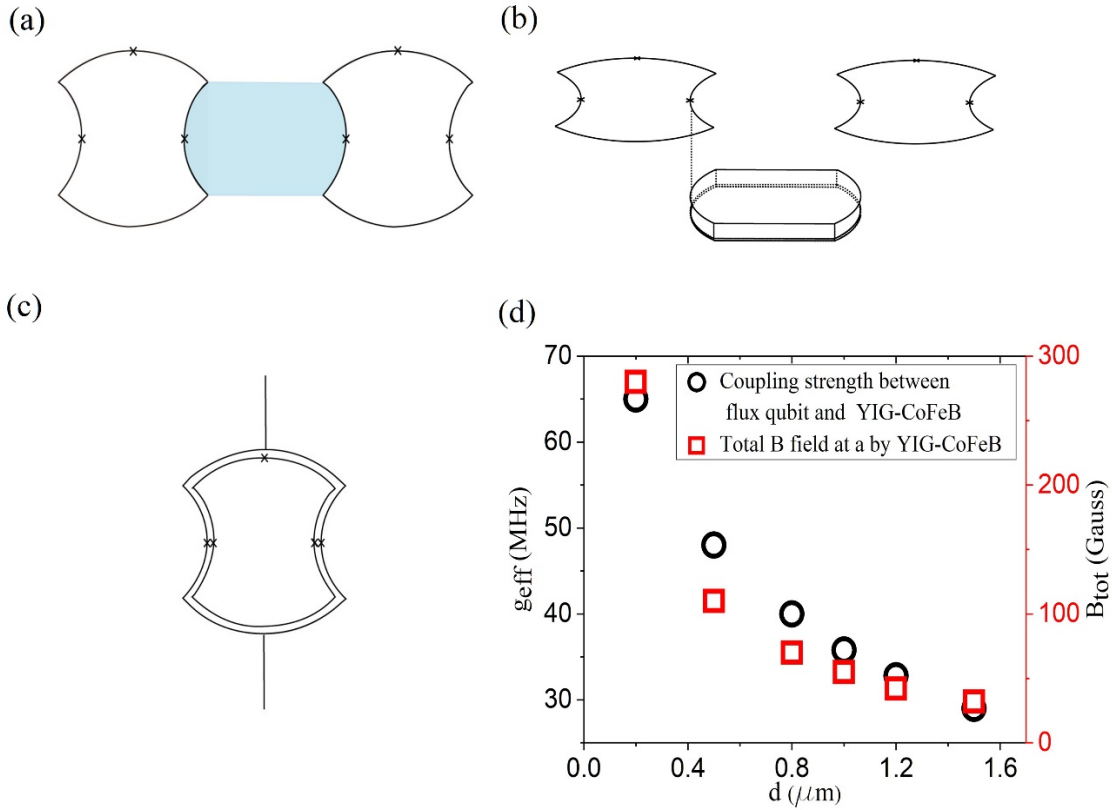

Fig. 4 **Proposed setup for a tunable switch between two shape-modified flux qubits utilizing (with) YIG thin film.** (a) two shape modified flux qubits are placed at a distance d above the 80 nm thick YIG thin film, which is capped with 10 nm CoFeB layer on one side. Special geometry of flux qubits is to decrease mutual inductance and detail dimensions of both flux qubits and YIG thin film are given in the context. (a) the sideview (b) the top view. (c) a special designed squid loop used for reading out the state of flux qubit. Mutual inductance between flux qubit and squid loop is given in the context and reading out one flux qubit will not be influenced much by the neighboring qubit. (d) the absolute value of effective coupling strength (left axis) between one flux qubit and YIG thin film and the total magnetic field (right axis) at point p as in (a) created by YIG thin film as a distance of d.

As demonstrated above, different from coupling to spin ensembles or Kittel mode of spin waves, the coupling of the flux qubit with finite-wavelength (fundamental) spin wave mode has an extra phase term, which enables us to obtain the coupling strength and proposed a scheme to hybridize flux qubit with a perpendicularly standing spin wave in the YIG thin film. We further show the PSSW spin wave mode in an YIG thin film can switch "on" and "off" the coupling between two flux qubits and generate entanglement. Our results manifest that it is possible to couple flux qubits while suppressing cross-talk. This opens a possibility of utilizing YIG thin film for quantum information processing.

The authors thank Huaiyang Yuan, Peihao Huang, Xiuhao Deng for fruitful discussions. This work is supported by Key-Area Research and Development Program of GuangDong Province (No. 2018B030326001), the National Key Research and Development Program of China (2016YFA0300802), the National Natural Science Foundation of China (Grants No. 11704022, No. U1801661), the Guangdong Innovative and Entrepreneurial Research Team Program (Grant No. 2016ZT06D348), the Science, Technology and Innovation Commission of Shenzhen Municipality (Grant No. KYTDPT20181011104202253).